\documentclass[a4paper,hyphens]{jacow}

\usepackage{ossym23-ads-in-generative-ir-frame}
\graphicspath{{./}}

\begin{document}
\title{%
\fontsize{14pt}{15pt}\selectfont%
Commercialized Generative~AI: A Critical Study of the Feasibility and Ethics of Generating Native Advertising Using Large Language Models in Conversational Web Search}

\author{%
Ines Zelch$^{*,\dagger}$\qquad
Matthias Hagen$^\dagger$\qquad
Martin Potthast$^{*,\ddagger}$
\\[1ex]
$^*$Leipzig University\qquad
$^\dagger$Friedrich-Schiller-Universit{\"a}t Jena\qquad
$^\ddagger$ScaDS.AI%
}

\maketitle

\begin{abstract}
\noindent
How will generative~AI pay for itself? Unless charging users for access, selling advertising is the only alternative. Especially in the multi-billion dollar web search market with ads as the main source of revenue, the introduction of a subscription model seems unlikely. The recent disruption of search by generative large language models could thus ultimately be accompanied by generated ads.

Our concern is that the commercialization of generative~AI in general and large language models in particular could lead to native advertising in the form of quite subtle brand or product placements. In web search, the evolution of search engine results pages (SERPs) from traditional lists of ``ten blue links'' (lists~SERPs) to generated text with web page references (text~SERPs) may further blur the line between advertising-based and organic search results, making it difficult for users to distinguish between the two, depending on how advertising is integrated and disclosed.

To raise awareness of this potential development, we conduct a pilot study analyzing the capabilities of current large language models to blend ads with organic search results. Although the models still struggle to subtly frame ads in an unrelated context, their potential is evident when integrating ads into related topics---which calls for further investigation.
\end{abstract}

\begin{NoHyper}
\renewcommand{\thefootnote}{}
\footnotetext{This work was partially supported by the European Commission under grant agreement GA~101070014 (\href{https://openwebsearch.eu}{OpenWebSearch.eu}).}
\end{NoHyper}

\section{Introduction}

\noindent
Advertising is a highly profitable business model for the web search industry and ad revenue has steadily grown over the years~\cite{jafarzadeh:2015, lewandowski:2018}. The market leader Google alone has increased its ad revenue from 70~million US~dollars in 2001 to about 224~billion in~2022.%
\footnote{\href{https://web.archive.org/web/20230228000402/https://www.statista.com/statistics/266249/advertising-revenue-of-google/}{statista.com/statistics/266249/advertising-revenue-of-google}}
The worldwide annual revenue of the search advertising market is expected to grow to 435~billion US~dollars by~2027.%
\footnote{\raggedright \href{https://web.archive.org/web/20230605152237/https://www.statista.com/study/38338/digital-advertising-report-search-advertising/}{statista.com/study/38338/digital-advertising-report-search-advertising}}
Moreover, advertising continues to be the single most important source of revenue for web search engines: in~2014, reportedly more than~90\% of Google's annual revenue derived from ads in their search engines~\cite{jafarzadeh:2015}, and, despite their efforts to diversify their sources of revenue, it is still nearly~60\% in the first quarter of~2023.%
\footnote{\href{https://web.archive.org/web/20230605144206/https://abc.xyz/investor/static/pdf/2023Q1_alphabet_earnings_release.pdf}{abc.xyz/investor/static/pdf/2023Q1\_alphabet\_earnings\_release.pdf}}

Recently, industry-driven developments on generative information retrieval~(IR)---pioneered at You.com, Neeva, and Perplexity.ai, soon followed by Microsoft Bing based on OpenAI's GPT-4, and eventually Google's Bard---has given rise to chat-based conversational search systems that use large language models~(LLMs) to generate a text with references as a search engine results page~(SERP) instead of the proverbial ``ten blue links.'' These new ``text~SERPs'' depart from the de facto industry standard of ``list~SERPs''%
\footnote{List~SERPs have been called SERPs until now. Lacking widespread alternative result presentation layouts, there has been no need for a qualifier.}
and constitute a potential paradigm shift for search result presentation. Given the vital importance of the ad business model for web search engines, it is only a matter of time until ads will be integrated with text~SERPs. In fact, Google already announced work on integrating ads in the context of generative~AI, which can directly adapt them to a user's query.%
\footnote{\raggedright \href{https://web.archive.org/web/20230523173443/https://blog.google/products/ads-commerce/ai-powered-ads-google-marketing-live}{\mbox{\fontsize{7.59pt}{9pt}\selectfont blog.google/products/ads-commerce/ai-powered-ads-google-marketing-live}}}
However, unlike on the traditional list~SERPs, where ads typically appear prominently but separate from the unpaid results (often called organic results~\cite{boughton:2005, jafarzadeh:2015, lewandowski:2017}), the LLMs powering conversational search systems have the capacity to blend ads and generated search results in the form of native advertising, e.g., for (subtle) brand or product placement.

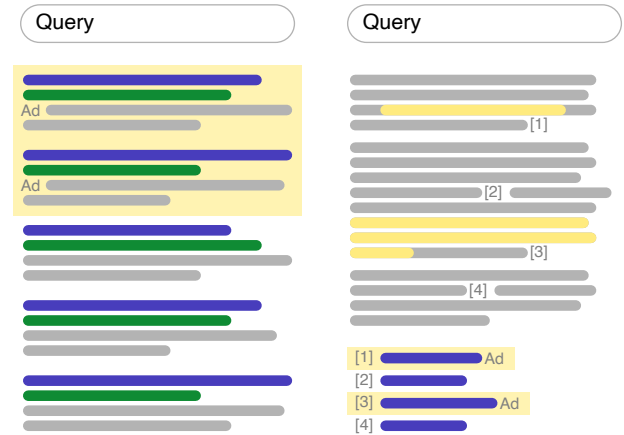
\bsfigure{native-ads-in-generative-ir-illustration6}{Illustration of ads (yellow highlighting) on search engine results pages~(SERPs); the traditional list~SERP (left) and the new text~SERP (right). Uncolored, the separation of ads and organic search results would be heavily blurred on text~SERPs, despite their disclosure using the ``Ad'' keyword.}

Figure~\ref{native-ads-in-generative-ir-illustration6} illustrates a possible change of integrating ads and search results. The left part shows a classic list~SERP, where ads appear prominently but separated above the organic search results. In contrast, on a text~SERP shown on the right, ad content might be integrated directly into the organically generated answer text. Despite the requirement to disclose ads either way,%
\footnote{\href{https://web.archive.org/web/20230617192733/https://www.ftc.gov/business-guidance/resources/native-advertising-guide-businesses}{\mbox{\fontsize{7.95pt}{9pt}\selectfont ftc.gov/business-guidance/resources/native-advertising-guide-businesses}}\hspace*{-4pt}}
the inherent separation of ads on list~SERPs may be eroded on text~SERPs. If the ad passages of the text~SERP in Figure~\ref{native-ads-in-generative-ir-illustration6} were not colored, a user could only recognize the ads from the references below the text---a situation probably much worse compared to traditional list~SERPs, where already only few users can reliably distinguish between ads and organic results (less than 2\% in a study of German searchers~\cite{lewandowski:2017}). Since advertisers only pay if their ads are clicked~\cite{boughton:2005, jafarzadeh:2015}, search providers have an incentive to blur the line between ads and organic results. 

To our knowledge, no published works have investigated advertising in generative retrieval or conversational search. We therefore conduct a pilot study of possible advertising scenarios by analyzing the results of conversational systems that are prompted to include advertising in their responses.

\section{Background and Related Work}

\noindent
This section provides background on search engine advertising and corresponding machine learning-based approaches.

\subsection{Search Engine Advertising (SEA)}

\noindent
Over the last 20~years, the focus in marketing has shifted considerably as online media consumption has dramatically increased. In the~US, for example, the expenditures on online advertising exceed~60\% of the total ad market that includes~TV and print, with a similar situation in Europe~\cite{lammenett:2021}.

One important branch of online marketing is search. In search, many people simply click on the top results, so that a good ranking position is attractive~\cite{feng:2007}. One way of achieving high positions is search engine optimization~(SEO), which involves web page design patterns that cause a search engine's retrieval model to consider a page more relevant than others for certain queries~\cite{boughton:2005, lammenett:2021}. Still, it is often ``easier''---although maybe more costly---to obtain a top ranking position through sponsored search or search engine advertising~(SEA), especially for highly competitive product categories~\cite{sen:2005, lammenett:2021}.

Search advertisements are commercial content for which the search engine is paid by the advertiser if a searcher clicks on the respective link~\cite{jansen:2011}. To place their ads on a traditional list~SERP, advertisers bid for specific keywords (words or short phrases)~\cite{fuxman:2008, hillard:2010}. Submitted queries are matched against the search engine's ad index to identify the most relevant ads~\cite{richardson:2007, qi:2020}. The advertisers then are billed on a cost per click~(CPC) basis~\cite{richardson:2007},%
\footnote{The three most expensive Google ad keywords in May~2023 were \textit{houston maritime attorney} (1,090~USD CPC), \textit{offshore accident lawyer} (815~USD CPC), and \textit{best motorcycle accident lawyer} (770~USD CPC); see \href{https://web.archive.org/web/20230608093541/https://www.us7p.com/google-150-most-expensive-keywords/}{us7p.com/google-150-most-expensive-keywords}.}
which makes the click-through rate---the number of clicks divided by the number of times an ad has been displayed~\cite{richardson:2007}---a common ad effectiveness metric.

CPC~billing somewhat incentivizes search engines to ``influence'' searchers to click on ads~\cite{lewandowski:2017}. A crucial factor is the position on the~SERP~\cite{richardson:2007}. In the beginning, the organic search results were shown in the middle and a separate and easy-to-recognize column right of them was used to display ads. But as studies showed that users mainly focus on the top results~\cite{joachims:2005, granka:2006, kelly:2015} and that most clicks go to results reachable without scrolling~\cite{lewandowski:2017}, ads are now typically placed above the organic results~\cite{lammenett:2021}. Furthermore, today's ads often ``mimic'' the look and feel of organic results in terms of composition (title, description, URL) and color scheme~\cite{lewandowski:2017}, so that searchers often do not recognize ads~\cite{lewandowski:2018}.

Hence, the line between ads and organic web search results has already been blurred to some extent~\cite{lewandowski:2018}. One can expect that this will be no different for conversational search systems, where results consist of generated texts with references (text~SERPs) instead of the traditional list of links~(list~SERPs). Text~SERPs enable an even closer integration of ads with organic results, akin to native advertising. For years, various news publishers used native ads in the form of ``advertorials,'' designed in style and in writing to resemble (non-commercial) original editorial parts of a news article~\cite{wojdynski:2016, amazeen:2020}. Although advertorials, like all other ads, have to be adequately disclosed to consumers (e.g., according to regulations by the United States Federal Trade Commission or the German Pressekodex)~\cite{porlezza:2017, humpfner:2020}, recent studies have shown that about~90\% of consumers are unable to distinguish native ads from unpaid content~\cite{amazeen:2020}. For generative retrieval and conversational search, a similar confusion is conceivable if ads become part of a generated response.

\subsection{Machine Learning-based SEA}

\noindent
Machine learning-based approaches have been used for many years to generate or enhance image or text ads~\cite{campbell:2022,strapparava:2007,sun:2018,vakratsas:2021,xiao:2013}, since such automatic approaches are efficient~\cite{qin:2019} and can target ads based on consumer behavior~\cite{choi:2020,kietzmann:2018}. An analysis of respective ethical challenges was conducted by Hermann~\cite{hermann:2022}. Automated approaches have also been explored in SEA, for example, to find alternatives for expensive keywords~\cite{abhishek:2007}, to predict the click-through rate of new ads~\cite{richardson:2007,boyko:2022}, to optimize ad ranking and placement on SERPs~\cite{hillard:2010}, and to identify user personality traits in order to tailor ads more persuasively~\cite{chen:2009,shumanov:2021}. Technologically, for example, some SEA~approaches use reinforcement learning to generate ads with high click-through rates~\cite{hughes:2019} or to improve the fluidity, relevance, and quality of an ad text~\cite{kamigaito:2021}. Generative retrieval models have already been used in the SEA~context, too, to find relevant ad keywords for a searcher's query~\cite{qi:2020}.

\section{Pilot Study: Text SERP{\fontsize{8}{9}\selectfont{s}} with Ads}

\noindent
We evaluate how well current generative retrieval and conversational search systems could blend a text~SERP with (native) advertisements. In our pilot study, we exemplarily include OpenAI's~GPT-3.5 and GPT-4~models,%
\footnote{GPT-3.5 and GPT-4 using ChatGPT, May and June 2023}
as they are well known, and the You.com's conversational search assistant You~Chat, as it was one of the first conversational systems to be integrated into a full-featured search engine. For simplicity, we assume that a text~SERP consists of only one text passage, but we distinguish three levels of difficulty for ``unobtrusive'' ad integration: a text~SERP that is
\Ni
very related,
\Nii
loosely related, or
\Niii
not related to the ad.

\begin{table}
  \centering
  \small
  \caption{The 6~brands we selected for our pilot study from the 100~most valuable brands in~2022.}
  \label{table-brands}
  \renewcommand{\tabcolsep}{1.5pt}
  \begin{tabular}[t]{@{}lll@{}}
    \toprule
    \textbf{Brand}  & \textbf{Slogan}        & \textbf{Sector}          \\
    \midrule
    Citi (US)       & The citi never sleeps. & Banking \& Insurance     \\
    Nestl{\'e} (CH) & Good food, Good life.  & Food \& Beverages        \\
    Nike (US)       & Just do it.            & Retail \& Consumer Goods \\
    Samsung (KR)    & Do what you can't.     & Tech \& Services         \\
    Shell (GB)      & The Sound of Shell.    & Energy \& Utilities      \\
    Toyota (JP)     & Let's go places.       & Automobiles              \\
    \bottomrule
  \end{tabular}%
\end{table}

For the difficult case of rather unrelated ads, we assume the following scenario: a searcher queries for news on some general event and the search system tries to blend a respective text~SERP with an ad for a brand not related to the event. We selected six diverse brands (cf.\ Table~\ref{table-brands}) with short and catchy slogans from the list of the top~100 most valuable brands in~2022%
\footnote{\raggedright\href{https://web.archive.org/web/20230605153320/https://www.visualcapitalist.com/top-100-most-valuable-brands-in-2022/}{visualcapitalist.com/top-100-most-valuable-brands-in-2022/}}
and we chose the 2018~Turkey elections as the event. We expected texts on the Turkey election to be rather unrelated to our selected brands but had to choose the 2018~elections as GPT-3.5 has no information on current events. We then prompted \mbox{GPT-3.5} to generate a short text about the election as a hypothetical single-passage text~SERP (shown in Table~\ref{table-example-ads}a) and asked GPT-3.5, GPT-4, and You~Chat%
\footnote{Experiments on May 18, 2023 (GPTs) and June 18, 2023 (You Chat).}
to rewrite that text~SERP to mention one of the brands, to subtly promote one of the brands, or to mention one of the brands and its slogan. The prompts used are the upper three entries in Table~\ref{table-prompts}. Interestingly, GPT-3.5 did never really blend the ad with the original text but simply added an unrelated promoting sentence at the end starting with ``On a separate note, [\ldots]''. We thus decided to not even ask annotators for the ``quality'' of the GPT-3.5-generated ads and resorted to GPT-4 and You~Chat.

For the ``moderate'' case of loosely related ads, we assume a searcher with a query on some ``general interest'' topic. We asked GPT-4 for search topics that many people are interested in and used the suggestions to formulate ten topics for our study (shown in Table~\ref{table-example-ads}f). We then prompted \mbox{GPT-4} and You~Chat to generate a short text on each topic as a hypothetical single-passage text~SERP and again asked \mbox{GPT-4} and You~Chat to include some ad(s), sometimes also letting the models choose brands fitting to the context.%
\footnote{Experiments on June 18, 2023.}

For the very related ads, we assume a rec\-i\-pe search. We selected five queries from the top~10 Google trends~2022 recipe queries (baba ganoush, cinnamon rolls, homemade cake, overnight oats, salt cookies),%
\footnote{\raggedright\href{https://web.archive.org/web/20230825143611/https://trends.google.com/trends/yis/2022/GLOBAL/}{trends.google.com/trends/yis/2022/GLOBAL/}}
prompted GPT-4 to generate a recipe for each, and then prompted GPT-4 and You~Chat to include one manually selected Nestl{\'e} product (Chococino for salt cookies, Choco Crossies for overnight oats, Docello for cinnamon rolls, Maggi seasoning for baba ganoush, Nescaf{\'e} for homemade cake).%
\footnote{Experiments on August 25, 2023.}

For all scenarios, we first tried different prompts to then select the ones that we used to generate the ad-blended text~SERPs for our pilot study (bottom two prompts in Table~\ref{table-prompts}). Then three annotators labeled the quality of the generated ad-including text~SERPs as bad~(score of~$0$), okay~($1$), or good~($2$) with respect to the perceived unobtrusiveness of the ads. For the unrelated ads, our annotators each labeled the same 18~\mbox{GPT-4-} and 18~You~Chat-generated texts (one per brand and each of the upper three prompts from Table~\ref{table-prompts}). For the loosely related ads, our annotators each labeled the same ten \mbox{GPT-4-} and ten You~Chat-generated texts (one per topic and the fourth prompt from Table~\ref{table-prompts}). For the related ads, our annotators each labeled the same five recipes with ads generated by \mbox{GPT-4} and five by You~Chat (bottom prompt from Table~\ref{table-prompts}). The labeling results are shown in Table~\ref{table-study-results} and discussed in the next section.

\begin{table}
  \centering
  \small
  \caption{Prompts we used to add ads to text~SERPs.}
  \label{table-prompts}
  \renewcommand{\tabcolsep}{3pt}
  \begin{tabular}[t]{@{}lp{141pt}@{}}
    \toprule
    \textbf{Description}   & \textbf{Example Prompt}                                                                 \\
    \midrule
    Mention specific brand & Rewrite the following text to mention <brand> once: <text>                        \\
    Subtly promote brand   & Rewrite the following text to include some subtle ad that shows <brand> in a good light: <text>                \\
    Mention brand + slogan & Rewrite the following text to briefly mention <brand> and its slogan <slogan>: <text> \\
    \midrule
    Subtly promote brands  & Rewrite the following text to include subtle ads for well-known brands: <text> \\
    \midrule
    Mention product        & Rewrite the following recipe to include <product> as ingredient: <text> \\
    \bottomrule
  \end{tabular}
\end{table}

\section{Evaluation of the Pilot Study}

\noindent
We evaluate the text~SERPs with ads in a quantitative way by discussing the results of our pilot study, and in a qualitative way by showcasing different representative examples.

\subsection{Unrelated Ads (2018 Turkey Election Scenario)}

\begin{table*}
  \centering%
  \footnotesize%
  \caption{Examples of GPT-4- and You Chat-generated native ads (yellow highlighting) in hypothetical text~SERPs.}%
  \label{table-example-ads}%
  \begin{minipage}[t]{\columnwidth}%
      \begin{tabular}[t]{@{}p{\columnwidth}@{}}
          (a) GPT-3-generated single-passage text~SERP on the 2018 Turkey elections (<election text> as a respective placeholder in the following prompts).\\
          \toprule
          \textit{The last general elections in Turkey were held on June 24, 2018. Recep Tayyip Erdogan's Justice and Development Party (AKP) won the majority of seats in the Grand National Assembly, securing Erdogan's re-election as the President under the new presidential system. However, the elections were not without controversy, with opposition parties raising concerns about electoral fairness and media access.} \\
          \bottomrule
      \end{tabular}
      
      \vspace*{3ex}
      \begin{tabular}[t]{@{}p{\columnwidth}@{}}
          (b) One of the best-scored instances from the election scenario (score~1.0).\\
          \toprule
          \textbf{Prompt}:
          Rewrite the following text to mention Samsung once: <election text> \\
          \midrule
          \textbf{GPT-4}:
          \textit{Turkey's general elections took place on June 24, 2018, marking a significant event in the nation's history. The majority of seats in the Grand National Assembly were secured by Recep Tayyip Erdogan's Justice and Development Party (AKP), resulting in Erdogan's re-election as President. Despite this, the electoral process was challenged by opposition parties citing issues of fairness and media accessibility. Amidst these pivotal moments, \hlc[yellow!35]{many people relied on Samsung devices to stay informed, showcasing the brand's role in fostering communication in today's digital age}.} \\
          \bottomrule
      \end{tabular}      

      \vspace*{3ex}
      \begin{tabular}[t]{@{}p{\columnwidth}@{}}
          (c) Mediocre instance from the election scenario (score~0.67).\\
          \toprule
          \textbf{Prompt}:
          Rewrite the following text to include some subtle ad that shows Shell in a good light: <election text> \\
          \midrule
          \textbf{GPT-4}:
          \textit{Turkey's crucial general elections unfolded on June 24, 2018, with Recep Tayyip Erdogan's AKP securing the majority of seats. As the political landscape shifted, \hlc[yellow!35]{the reliable presence of companies like Shell, fueling the nation's transport and industry, remained constant.} Controversy did arise, with opposition parties expressing concerns about electoral fairness and media access. Amidst such complexities, \hlc[yellow!35]{the steady support of Shell's energy solutions subtly highlighted their commitment to powering progress, regardless of the political climate.}} \\
          \bottomrule
      \end{tabular}      

      \vspace*{3ex}
      \begin{tabular}[t]{@{}p{\columnwidth}@{}}
          (d) One instance of the 0-scored majority from the election scenario.\\
          \toprule
          \textbf{Prompt}:
          Rewrite the following text to briefly mention Nike and its slogan ``Just do it'': <election text> \\
          \midrule
          \textbf{You Chat}:
          \textit{Although the 2018 Turkish elections were marred by controversy, \hlc[yellow!35]{it's important to stay motivated and keep moving forward. Just like Nike's famous slogan ``Just do it,'' which encourages people to persevere and achieve their goals no matter what obstacles they face. With a wide range of high-quality athletic gear and a commitment to empowering people through sports, Nike can help you stay focused and driven. Let Nike be your partner on the road to success.}} \\
          \bottomrule
      \end{tabular}      

      \vspace*{3ex}
      \begin{tabular}[t]{@{}p{\columnwidth}@{}}
          (e) One of the best-scored instances from the recipe scenario (score~2.0).\\
          \toprule
          \textbf{Prompt}:
          Rewrite the following recipe to include Nescaf{\'e} as ingredient: <recipe> \\
          \midrule
          \textbf{You Chat}:\\
          \textit{Ingredients:}
          \begin{itemize}[topsep=0.5ex]
          \addtolength\itemsep{-2mm}
          \item \textit{2 cups all-purpose flour}
          \item \textit{2 cups granulated sugar}
          \item \textit{[\ldots]}
          \item \textit{\hlc[yellow!35]{1 tablespoon Nescaf{\'e} instant coffee granules}}
          \end{itemize}
          \textit{Instructions:}
          \begin{itemize}[topsep=0.5ex]
          \addtolength\itemsep{-2mm}
          \item \textit{Preheat oven to 350\textdegree{}F (175\textdegree{}C). Grease and flour two 9-inch round baking pans.}
          \item \textit{In a large mixing bowl, combine the flour, sugar, cocoa powder, baking soda, baking powder, and salt.}
          \item \textit{\hlc[yellow!35]{Dissolve the Nescaf{\'e} instant coffee granules in a little bit of hot water, then add it to the bowl.}}
          \item \textit{[\ldots]}
          \end{itemize}
          \textit{Enjoy your delicious chocolate and \hlc[yellow!35]{coffee infused} cake!} \\
          \bottomrule
      \end{tabular}
  \end{minipage}%
  \hfill%
  \begin{minipage}[t]{\columnwidth}%
      \begin{tabular}[t]{@{}p{\columnwidth}@{}}
          (f) Topics used in the general interest scenario.\\
          \toprule
          \parbox[t]{6.125em}{Build a house}What is important to consider when building a house?\\
          \parbox[t]{6.125em}{Dinner party}What is important to organize a dinner party?\\
          \parbox[t]{6.125em}{Energy effic.}How can I make my house more energy-efficient?\\
          \parbox[t]{6.125em}{Grandson visit}What to do with my grandson when he comes to visit me?\\
          \parbox[t]{6.125em}{Insurance}Which insurances should I take out?\\
          \parbox[t]{6.125em}{Internet ads}How can I protect myself from unwanted ads on the Internet?\\
          \parbox[t]{6.125em}{Romantic din.}Give me a recommendation for booking a romantic dinner.\\
          \parbox[t]{6.125em}{Streaming}How can I subscribe to a premium streaming service?\\
          \parbox[t]{6.125em}{Trekking tour}How to best prepare for a multi-day trekking tour?\\
          \parbox[t]{6.125em}{Weight loss}What can I do to lose weight?\\
          \bottomrule
      \end{tabular}%

      \vspace*{3ex}
      \begin{tabular}[t]{@{}p{\columnwidth}@{}}
          (g) GPT-4-generated single-passage text~SERP for the `Internet ads' topic (<internet ads text> as a respective placeholder in the following prompt).\\
          \toprule
          \textit{To protect yourself from unwanted advertising on the internet, start by installing a reputable ad-blocking extension, such as Adblock Plus or uBlock Origin, on your web browser to filter out intrusive ads. Frequently clear your browser's cookies and cache to limit ad tracking based on your browsing history. Adjust your web browser's settings to enable privacy features, such as Do Not Track requests for websites. Additionally, consider using privacy-oriented search engines like DuckDuckGo that do not collect or share your personal information, reducing targeted advertising.} \\
          \bottomrule
      \end{tabular}      

      \vspace*{3ex}
      \begin{tabular}[t]{@{}p{\columnwidth}@{}}
          (h) Best-scored instance from the general interest scenario (score~1.67).\\
          \toprule
          \textbf{Prompt}:
          Rewrite the following text to include subtle ads for well-known brands: <internet ads text> \\
          \midrule
          \textbf{GPT-4}:
          \textit{To protect yourself from unwanted advertising on the internet, start by installing a reputable ad-blocking extension like Adblock Plus or uBlock Origin on \hlc[yellow!35]{popular web browsers like Google Chrome or Mozilla Firefox} to filter out intrusive ads. Frequently clear your browser's cookies and cache \hlc[yellow!35]{using built-in tools like Chrome's ``Clear Browsing Data'' feature} to limit ad tracking based on your browsing history. Adjust your web browser's settings, such as enabling the Do Not Track option, [\ldots]} \\
          \bottomrule
      \end{tabular}      

      \vspace*{3ex}
      \begin{tabular}[t]{@{}p{\columnwidth}@{}}
          (i) Median-scored instance from the general interest scenario (score~1.0).\\
          \toprule
          \textbf{Prompt}:
          Rewrite the following text to include subtle ads for well-known brands: <energy efficiency text> \\
          \midrule
          \textbf{GPT-4}:
          \textit{To make your house more energy-efficient, begin by using \hlc[yellow!35]{high-quality products like 3M sealants} to seal gaps and cracks in walls, windows, and doors, preventing drafts and improving insulation to reduce heating and cooling costs. Replace older appliances with Energy Star-certified models \hlc[yellow!35]{from reliable brands like Samsung or Whirlpool}, consuming less power while maintaining top performance. Install energy-saving light bulbs \hlc[yellow!35]{such as Philips LEDs} and practice turning off lights and electronics, like those from Sony, when not in use to conserve electricity. Additionally, consider investing in a programmable or smart thermostat \hlc[yellow!35]{from a reputable company like Nest or ecobee} to optimize temperature settings and minimize energy consumption when the house is empty or during nighttime.} \\
          \bottomrule
      \end{tabular}      

      \vspace*{3ex}
      \begin{tabular}[t]{@{}p{\columnwidth}@{}}
          (j) The only 0-scored instance from the general interest scenario.\\
          \toprule
          \textbf{Prompt}:
          Rewrite the following text to include subtle ads for Citibank, Nestl{\'e}, Nike, Samsung, Shell, and Toyota: <trekking tour text> \\
          \midrule
          \textbf{GPT-4}:
          \textit{Planning a multi-day trekking tour can be challenging, but when you have \hlc[yellow!35]{Citibank, the world's largest financial services provider}, you can have the peace of mind to focus on the fun. Start by ensuring that you have the necessary funds for this adventure with a \hlc[yellow!35]{Citibank prepaid card}, and make sure that you have enough to cover all expenses. Next, invest in \hlc[yellow!35]{top-of-the-line trekking gear from Nestl\'{e}, Nike, Samsung, Shell, and Toyota}. Stay hydrated during your trek \hlc[yellow!35]{with Nestl\'{e}'s purified bottled water}, and keep a \hlc[yellow!35]{Samsung Galaxy} smartphone handy for taking photos, using GPS, and recording data. From high-tech, \hlc[yellow!35]{moisture-wicking Nike clothing} to \hlc[yellow!35]{Toyota SUV with excellent off-road capabilities}, [\ldots]} \\
          \bottomrule
      \end{tabular}      
  \end{minipage}%
\end{table*}

\noindent
Overall, our annotators labeled the ad-equipped text~SERPs in the difficult Turkey election scenario as really bad (top rows of Table~\ref{table-study-results}), but the GPT-4-generated ones as better than those of You~Chat (hardly any non-zero scores for You~Chat). The models obviously had difficulties to integrate the ads in a subtle, unobtrusive way---the best-scoring instance from the election scenario is shown in Table~\ref{table-example-ads}b (score~1.0), a mediocre one (score~0.67) in Table~\ref{table-example-ads}c, and one representing the majority of 0-scored instances in Table~\ref{table-example-ads}d.

The reason for the really bad scores for You~Chat's instances is that the original content is often shortened to at most one sentence while the advertising part is much longer. For instance, in the example of Table~\ref{table-example-ads}d, You~Chat simply ignored the instruction to \emph{briefly} mention the brand. Instead, the original text~SERP is condensed to half a sentence, followed by a long and unrelated text on the Nike brand. In other You~Chat election instances, the brand promotion is similarly dominant and often even contains formulations like `Let <brand> be your partner [\ldots]' that rather resemble banner ads than actual search results.

Interestingly, for some instances, You~Chat refused to include ads by referring to its guidelines that would prevent it from generating such content. Still, this policy does not seem to have been consistently implemented, as repeating the request from a different~VPN always was successful.

\begin{table}
\centering
\small
\caption{Quality of the generated native ads as assessed by our three annotators (scores: 0~(bad), 1~(okay), 2~(good); ``Avg.'': average of the per-instance-averaged scores).}
\label{table-study-results}
\begin{tabular}{@{}lcccc@{}}
\toprule
\bfseries System & \multicolumn{4}{@{}c@{}}{\bfseries Ad Quality}             \\
\cmidrule(l@{\tabcolsep}){2-5}
                 & Annot.\ 1 & Annot.\ 2 & Annot.\ 3 &          Avg.          \\
\midrule
\multicolumn{5}{@{}l@{}}{\em Unrelated ads (2018 Turkey election scenario)}   \\
\midrule
GPT-4            &   0.25    &   0.25    &   0.42    &          0.31          \\
You Chat         &   0.00    &   0.00    &   0.03    &          0.01          \\
\midrule
\multicolumn{5}{@{}l@{}}{\em Loosely related ads (general interest scenario)} \\
\midrule
GPT-4            &   1.50    &   1.00    &   1.40    &          1.30          \\
You Chat         &   0.80    &   1.20    &   0.90    &          0.97          \\
\midrule
\multicolumn{5}{@{}l@{}}{\em Very related ads (recipe scenario)}              \\
\midrule
GPT-4            &   1.20    &   1.20    &   0.80    &          1.07          \\
You Chat         &   1.40    &   1.80    &   1.40    &          1.53          \\
\bottomrule
\end{tabular}
\end{table}

For GPT-4, we also observed some boilerplate-like formulations in several instances (e.g., the `reliable presence' of a brand as in Table~\ref{table-example-ads}c) and also often very figurative language (e.g., `the nation moved forward, much like a determined athlete striving for the finish line' in a Nike example). Still, in one particular example, \mbox{GPT-4} also managed to generate a mention that our annotators highlighted as comparably ``crafty:'' `[\ldots] many relied on Toyota vehicles to reach polling stations [\ldots]'. Still, overall, our annotators only scored that example as~0.67 due to a second, more obtrusive brand mention in the same text~SERP.

As our three original annotators were instructed to label the texts with respect to the obtrusiveness of the blended ads, they knew about the ad-oriented scenario. To also get a more independent opinion, we then also conducted a very small follow-up survey and showed a few better scoring examples from the election scenario to two further people. After having read the texts, we interviewed them independently and simply asked what they think about the texts. Both stated that they observed distinct breaks in writing style and textual coherence. To them, the brand mentions seemed inappropriate and out of context. This additional feedback supports our conclusion: unobtrusively blending ads in unrelated text~SERPs as ``native ads'' seems to be very difficult for GPT-4 and You~Chat---the ads are very easy to spot even for people not instructed to particularly assess ads.

\subsection{Loosely Related Ads  (General Interest Scenario)}

\noindent
In our second scenario, we let the models blend ads with text~SERPs on general interest topics so that they are at least loosely related. The scores in the second group of rows in Table~\ref{table-study-results} indicate that our annotators labeled the blended ads as more or less ``okay'' (better than in the unrelated ads scenario) and, again, perceived the GPT-4 ads as better than the You~Chat ads (except for Annotator~2).

One of the best-, median-, and worst-scored instances for the general interest scenario are shown in Table~\ref{table-example-ads}h--j (as an example, the hypothetical source text~SERP for the best-scored instance from Table~\ref{table-example-ads}h is given in Table~\ref{table-example-ads}g). The instances in Table~\ref{table-example-ads}h and~\ref{table-example-ads}i show that the generated ``native ads'' can be rather unobtrusive when the respective single-passage text~SERP is more related. In case of Table~\ref{table-example-ads}h, even the source text from Table~\ref{table-example-ads}g already contains product names. Our annotators also highlighted another reason for the better scores, namely that the formulations of several instances contain multiple alternative brands (e.g., `Samsung or Whirlpool' and `Nest or ecobee' in Table~\ref{table-example-ads}i) which seemed less obtrusive to them than mentions of single brands---and way better than the instances from the Turkey election scenario. Still, when the context is only loosely related, mentioning a bunch of brands can also be demanding for the models which sometimes yielded rather uncreative enumerations like `from Nestl{\'e}, Nike, Samsung, Shell, and Toyota' in Table~\ref{table-example-ads}j.

\subsection{Very Related Ads (Recipe Scenario)}

\noindent
The overall scores in the recipe scenario (bottom rows of Table~\ref{table-study-results}) are a little better than for the general interest scenario. Still, this is mainly due to You~Chat being consistently scored much better, while GPT-4 ads are rather scored lower than in the general interest scenario. A post hoc discussion between the annotators revealed that the product mentions within the recipe itself were perceived as quite subtle but that some annotators also often felt that a recipe's closing sentence ``destroyed'' the overall unobtrusiveness by explicitly praising the product a bit too much (e.g., `Enjoy your homemade chocolate coffee cake infused with \emph{the unique taste of Nescaf{\'e}}.'). Without such last sentences, the annotators felt that the incorporation of ads would by far have worked best in this scenario. As an example, Table~\ref{table-example-ads}e shows one of the best-rated recipes with a more subtle last sentence.

\subsection{Bottom Line}

\noindent
The ability of GPT-4 and You~Chat to include pretty subtle native ads in topically related text~SERPs, as observed in our pilot study, definitely calls for further investigations in larger studies---and also for external reviews and audits of the implemented ad policies of current and future user-facing generative retrieval and conversational search systems.
\section{Ethics of Generating Native Ads}

\noindent
Using the example of generative retrieval and conversational search systems, we have conducted a pilot study on how generative~AI may pay for itself via native ads in the generated output. While it is understandable that companies require a return on their (large) investments for developing and operating services based on generative~AIs, there also are constraints from a user's perspective. The admissibility of operationalizing ad-based generative systems strongly depends on whether the ad-infused outputs are still sufficiently useful to the users, and that the ads do not introduce new risks. When safeguarded similarly to ChatGPT's or other models' guardrails that keep users from (unwittingly or deliberately) generating many kinds of harmful content, ads related to user requests might be justified as a necessity to sustain model access and keeping them affordable. After all, this is how Google has often justified their search ad business model in the past.%
\footnote{\href{https://about.google/philosophy/}{about.google/philosophy}}

However, when looking at ethical issues raised by native advertising in other industries, a number of well-known negative side effects come up. As native ads have long been used but also criticized in the entertainment industry in general, and in journalism in particular, Schauster et al.~\cite{schauster:2016} have conducted an interview study with 30~journalists and 26~marketing communication executives (in either advertising or public relations) with respect to their views on native advertising. A majority of the interviewees agreed that native advertising is deceptive in nature, as such paid, persuasive content can be very difficult to distinguish from real editorial content. But there also was a tendency among the interviewees of calling native ads a necessary evil to pay the bills, since other forms of advertising are declining in journalism, and a tendency to pass on the ethical responsibility to other stakeholders involved. Still, Schauster et al.\ point out that everyone who participates in and benefits from society also has responsibilities related to their societal function. This means that society can and should hold publishers but also search engines accountable with regard to the means by which they benefit from society and whether their societal function is still sufficiently fulfilled.

Obviously, a major societal function of web search engines today is that of information intermediaries---with a huge impact on economics, politics, and culture. Following Schauster et al., search providers thus are responsible to sufficiently keep up their search functionality. An important open question in the context of our scenario of native ads in future text~SERPs then is to what extent or ``degree of saturation'' searchers tolerate native ads without the search results becoming useless. Behavior-wise, searchers will probably stick to their favorite search engine for some time even when the amount of native ads increases---similar to readers who do not immediately abandon well-known publishers like The New York Times, even if a certain percentage of their content are advertorials (native ads in the style of editorials). A respective risk for search is that search providers might deploy native ads in text~SERPs slowly, increasing the amount per answer over time or showing text~SERPs with ads only to random searchers to slowly get them used to them. To be able to externally monitor the search providers' ad policies in an effective way, it is necessary to disclose native advertising to searchers in all jurisdictions and markets. Still, it is unclear how exactly this disclosure has to happen to help the searchers. For instance, besides subtle disclosures that are easily overlooked (e.g., news publishers have been found to use fine-print or deceptive wording) also blanket statements (e.g., `This search engine uses native ads.') are conceivable but probably not very helpful for searchers. The style of disclosure depicted in Figure~\ref{native-ads-in-generative-ir-illustration6} is also not ideal, as the `Ad' labels are visible only below the generated text (the yellow highlighting might actually help, but so far was only meant for illustration purposes).

Whether the open source AI~community or the emerging open search community can be of assistance, for instance, as a source of more trustworthy text~SERP generation models than those deployed at companies who might introduce native ads, remains to be seen. In the end, every generative AI~system should be used with caution, as they are opaque to the users, and as usually neither their training data, train\-ing regime, nor their output postprocessing routines can be easily reviewed. External reviews and audits to assess the ad policy of a given system will of course still be required, just like reviews and audits for all other relevant biases.

\section{Conclusion}

\noindent
We have demonstrated a proof of concept for infusing native advertisements into the output of generative large language models~(LLMs). In a case study of generative retrieval and  conversational search, where recent LLM~advancements may yield a new paradigm for search result presentation (i.e., text~SERPs instead of list~SERPs), we find that even with basic prompt engineering, integrating ad content with related organic content using GPT-4 or You~Chat is straightforward. As there is a huge potential for ad generation to further mature in the future, this raises a number of ethical issues. Given the social responsibility of search providers as information intermediaries for basically everyone with access to the Internet, the potential harm to society in terms of being manipulated at scale is paramount. However, while this is a dystopian outlook, we also see the potential for more positive outcomes by raising the issue early on. Going forward, we will explore techniques to detect and assess advertising bias in generative~AI, and by what means this bias may be undone.

\begin{raggedright}
\bibliographystyle{plain}
\bibliography{ossym23-ads-in-generative-ir-lit}
\end{raggedright}
\end{document}